\documentclass[letterpaper,preprintnumbers,prd,twocolumn,nofootinbib,nobibnotes,showpacs]{revtex4}
\usepackage{amsfonts}
\usepackage{mathrsfs}
\usepackage{epsfig}
\usepackage{graphicx}%
\usepackage{dcolumn}
\usepackage{amsmath}

\makeatletter
\def\btt#1{\texttt{\@backslashchar#1}}%
\DeclareRobustCommand\bblash{\btt{\@backslashchar}}%
\makeatother
\begin{document}

\title{Modified gravity in Arnowitt-Deser-Misner formalism}
\author{Changjun Gao}\email{gaocj@bao.ac.cn}\affiliation{The
National Astronomical Observatories, Chinese Academy of  Sciences,
Beijing, 100012} \affiliation{{Kavli Institute for Theoretical
Physics China, CAS, Beijing 100190, China }}

\date{\today}

\begin{abstract}
Motivated by Ho$\check{\textrm{r}}$ava-Lifshitz gravity theory, we
propose and investigate two kinds of modified gravity theories, the $f(R)$ kind and the K-essence kind, in the
Arnowitt-Deser-Misner (ADM) formalism. The $f(R)$ kind
includes one ultraviolet (UV) term and one infrared (IR) term together with
the Einstein-Hilbert action. We find that these two terms naturally
present the ultraviolet and infrared modifications to the
Friedmann equation. The UV and IR modifications can avoid the past
Big-Bang singularity and the future Big-Rip singularity,
respectively. Furthermore, the IR modification can naturally account
for the current acceleration of the Universe.
The Lagrangian of K-essence kind modified gravity is
made up of the three dimensional
Ricci scalar and an arbitrary function of the extrinsic curvature
term. We find the cosmic
acceleration can also be naturally interpreted without invoking any kind
of dark energy. The static,
spherically symmetry and vacuum solutions of both theories are
Schwarzschild or Schwarzschild-de Sitter solution. Thus these modified gravity theories are viable for solar system
tests.
\end{abstract}
\maketitle

\section{Introduction}
Recently, Ho$\check{\textrm{r}}$ava
\cite{ho:20081,ho:20082,ho:20083,ho:20090} proposed a four
dimensional gravity theory in which the space and time are treated
on an unequal footing. The theory is very much interesting because
of its power counting renormalizability. Therefore one generally
believes that it is a ultraviolet (UV) complete of General
Relativity (GR). Up to now, much attentions have been paid to this theory
\cite{ho:20091,ho:20092,ho:20093,ho:20094,ho:20095,ho:20096,ho:20097,ho:20098,ho:20099,ho:200910,
ho:200911,ho:200912,ho:200914,ho:200915,ho:200916,ho:200917,ho:200918}.
The theory is formulated in the Arnowitt-Deser-Misner (ADM)
formalism \cite{adm:1962}. Motivated by this theory, we shall
present our modified gravities in the ADM formalism.

The four dimensional metric in the ADM formalism is given by

\begin{eqnarray}
 ds^2=-N^2dt^2+g_{ij}\left(dx^{i}+N^{i}dt\right)\left(dx^{j}+N^{j}dt\right)\;,
\end{eqnarray}
where $N,\ N_{i},\ g_{ij}$ are the lapse function, shift function
and the three dimensional metric, respectively. The Latin letters
$i,\ j$ runs over $1,\ 2,\ 3$. For a spacelike hypersurface with a fixed time, the extrinsic
curvature $K_{ij}$ is defined by

\begin{eqnarray}
 K_{ij}=\frac{1}{2N}\left(\dot{g}_{ij}-\nabla_i N_j-\nabla_j
 N_i\right)\;,
\end{eqnarray}
where dot denotes the derivative with respect to $t$.
In the ADM formalism, the four dimensional Ricci scalar can be decomposed as \cite{wald}
\begin{eqnarray}
R&=& R^{(3)}+K_{ij}K^{ij}-K^2\nonumber\\&&+2\nabla_{i}\left(n^{i}\nabla_{j} n^{j}\right)-2\nabla_{i}\left(n^{j}\nabla_{j}n^{i}\right)\;.
\end{eqnarray}
Here $n^{i}$ is the unit normal vector on the hypersurface and $R^{(3)}$ is the three dimensional Ricci scalar.
Rewrite the Hilbert-Einstein action
in the Einstein frame:
\begin{eqnarray}
 S=\int d^4x\sqrt{-g}\frac{R}{16\pi}\;,
\end{eqnarray}
in the ADM formalism:
\begin{eqnarray}
 S&=&\int dtd^3xN\sqrt{g^{(3)}}\frac{1}{16\pi}\left[R^{(3)}+K_{ij}K^{ij}-K^2\right.\nonumber\\&&\left.+2\nabla_{i}\left(n^{i}\nabla_{j} n^{j}\right)-2\nabla_{i}\left(n^{j}\nabla_{j}n^{i}\right)\right]\;,
\end{eqnarray}
where $g^{(3)}$ is the trace of three dimensional metric. One find the last two terms in the integrand
would contribute a boundary term which does not enter the equation of motion \cite{wald}. Therefore, the action
can be written as
\begin{eqnarray}
 S&=&\int dtd^3xN\sqrt{g^{(3)}}\frac{1}{16\pi}\left(R^{(3)}+K_{ij}K^{ij}-K^2\right)\;.
\end{eqnarray}
We stress that for nonlinear terms of Ricci scalar, $R^{n}$, the last two terms
would enter the equation of motion such that the $f(R)$ theory in the ADM formalism:
\begin{eqnarray}
 S&=&\int dtd^3xN\sqrt{g^{(3)}}\frac{1}{16\pi}f\left(R\right)+S_m\;,
\end{eqnarray}
is equivalent to that in the Jordan frame:
\begin{eqnarray}
 S=\int d^4x\sqrt{-g}\frac{1}{16\pi}f\left(R\right)+S_m\;.
\end{eqnarray}
However, if we neglect the boundary term for nonlinear Ricci scalar terms and take the modified gravity as follows
\begin{eqnarray}
S&=&\int dtd^3xN\sqrt{g^{(3)}}\frac{1}{16\pi}f\left(R^{(3)}+K_{ij}K^{ij}-K^2\right)\nonumber\\&&+S_m\;.
\end{eqnarray}
Then the theory would be different from the $f(R)$ version in the Jordan frame.
To make our theory different from the usual $f(R)$ version, we shall neglect the boundary terms in this paper.

Up to the lowest possible orders for IR and UV corrections, the modified gravity in the Einstein frame takes the form of
\begin{eqnarray}
 S=\int d^4x\sqrt{-g}\frac{1}{16\pi}\left(R+\alpha R^2+\frac{\beta}{R}\right)+S_m.
\end{eqnarray}
The theory has been investigated very extensively \cite{st:1977}. Correspondingly, we
will explore the first modified gravity theory in the ADM formalism:

\begin{eqnarray}
 S&=&\int dtd^3xN\sqrt{g^{(3)}}\frac{1}{16\pi}\left[\left(R^{(3)}+K_{ij}K^{ij}-K^2\right)
 \right.\nonumber
\\
&&\left.+\alpha\left(R^{(3)}+K_{ij}K^{ij}-K^2\right)^2\right.\nonumber
\\
&&\left.+\beta\left(R^{(3)}+K_{ij}K^{ij}-K^2\right)^{-1}\right]\;+S_m,
\end{eqnarray}
where $\alpha, \beta$ are two positive constants. We find that, in the ADM
formalism, the corresponding Friedmann equation is remarkably simple
and very different from that in the Jordan frame \cite{st:1977}.
Therefore, the theory will present a different cosmic evolution
history.

On the other hand, $K_{ij}K^{ij}-K^2$ may be understood as
a kinetic term of extrinsic curvature. Similar to the K-essence theory
\cite{k-essence:00}, we may construct another K-essence kind of modified gravity.
To this end, we define
\begin{eqnarray}
 X=K_{ij}K^{ij}-K^2\;,
\end{eqnarray}
then the second modified gravity we will explore can be
written as:

\begin{eqnarray}
 S&=&\int
 dtd^3xN\sqrt{g}\frac{1}{16\pi}\left[R^{(3)}+X+{F}\left(X\right)\right]
 +S_m,
\end{eqnarray}
where ${F}\left(X\right)$ is an arbitrary function of $X$. When
${F}\left(X\right)=const$, the theory reduces to General Relativity. Similar to the $f(R)$ modified gravity, we expect the nonlinear terms of
$X$ may arise in the quantum corrections to GR. With this modifications, we find the cosmic acceleration can also be interpreted without invoking any kind
of dark energy. It is interesting that this ``K-essence" can cross the phantom divide.

The paper is organized as follows. In Section II and Section IV, we investigate the
cosmological behavior of the $f(R)$ kind and the K-essence kind of modified gravity, respectively.  In Section III and Section V, we look for the
static, spherically symmetry and vacuum solutions.  In Section VI we
make the conclusion and discussion. Throughout the paper, we use the
units in which $c=G=\hbar=1$.
\section{Cosmology-f(R) kind}
Consider the spatially flat Friedmann-Robertson-Walker Universe
\begin{eqnarray}
 ds^2 = -N\left(t\right)^2dt^2+a\left(t\right)^2\left(dr^2+r^2d\Omega^2\right)\;,
\end{eqnarray}

So

\begin{eqnarray}
 K_{ij}=\frac{H}{N}g_{ij}\;,\ \ \ R^{(3)}_{ij}=0\;,
\end{eqnarray}

The action is given by
\begin{eqnarray}
\label{eq:fr}
 S=\int dtd^3x\frac{Na^3}{16\pi}\left(-\frac{6H^2}{N^2}+\frac{36\alpha H^4}{N^4}-\frac{\beta
 N^2}{6H^2}\right)+S_{m}\;.
\end{eqnarray}
Variation of the action with respect to $N$ and then put $N=1$, we
obtain the Friedmann equation

\begin{eqnarray}
\label{eq:fr}
 3H^2=8\pi\sum_i\rho_i+54\alpha H^4+\frac{\beta}{4H^2}\;,
\end{eqnarray}
where $\rho_i$ is the energy density for ith component of matters
which mainly include dark matter and radiation. We note that here
the Friedmann equation is remarkably simple and very different from
that in the Einstein frame \cite{st:1977}. Therefore, it will
present us a different cosmic evolution history. It is interesting that in many brane word models, the
modifications to Friedmann equation effectively corresponds to $H^4$ and $H^{-2}$ \cite{bin:2000,dva:2002,dva:2003}.

On the other hand,
variation of the metric with respect to $a\left(t\right)$, we obtain
the acceleration equation

\begin{eqnarray}
\label{eq:fr}
 2\dot{H}+3H^2&=&-8\pi \sum_i p_i+72\alpha H^2\dot{H}+54\alpha
 H^4\nonumber\\&&+\frac{\beta}{4}H^{-2}+\frac{\beta}{6}H^{-4}\dot{H}\;,
\end{eqnarray}
where $p_i$ is the pressure for the ith matter. We are able to
derive the energy conservation equation from the Friedmann equation
and the acceleration equation

\begin{eqnarray}
\label{eq:fr}
 \sum_i\left[\frac{d\rho_i}{dt}+3H\left(\rho_i+p_i\right)\right]=0\;.
\end{eqnarray}
If we assume there is no interaction between dark matter and
radiation, we will have
\begin{eqnarray}
\label{eq:fr}
 \frac{d\rho_i}{dt}+3H\left(\rho_i+p_i\right)=0\;.
\end{eqnarray}

So for convenience, we can only consider the Friedmann equation and
the energy conservation equitation. Put

\begin{eqnarray}
\label{eq:fr}
 \alpha=\frac{1}{192\pi \rho_U}\;,\ \ \ \beta=\frac{64\pi^2}{3}\rho_I^2\;,
\end{eqnarray}
where $\rho_U,\ \rho_I$ are constant energy densities. We assume $\rho_U$ is
on the order of Planck energy density, $\rho_U=\rho_p$. In order to explain the current
acceleration of the Universe, we find shortly later $\rho_I$ should
on the order of present-day cosmic energy density. Therefore they
represent the UV and IR modification of Friedmann equation. With this assumptions, we find the energy
density of $\alpha$ term is negligible for the present-day Universe:
\begin{eqnarray}
\frac{9}{32\pi\rho_U\rho_0}H^4|_{H=H_0}\simeq 10^{-123}\;.
\end{eqnarray}
This energy density becomes significantly only when the Hubble
radius is on the order of Planck length. Therefore, it is a UV
modification term.

For the $\beta$ term, We have
\begin{eqnarray}
\frac{16\pi^2\rho_I^2}{3H^2\rho_0}|_{H=H_0}\simeq
\textrm{O}\left(1\right)\;.
\end{eqnarray}
This term plays a great role in the present-day Universe. It is
negligible at very higher redshifts (large $H$) while becomes
significant in the future (small $H$). Therefore, it is an IR
modification.

\subsection{UV modification} In this subsection, we investigate the UV
modification. We find that the Big-Bang singularity can be safely
avoided. In the presence of only UV modification, the Friedmann equation is given by
\begin{eqnarray}
\label{eq:fr} 3H^2=8\pi {\rho}+\frac{9}{32\pi\rho_U}H^4\;.
\end{eqnarray}
It is a quadratic equation of $H^2$. Mathematically, we would have two roots for $H^2$. But physically, only one root
could reduce to the standard Friedmann equation in the limit of smaller $\rho$. We find the root takes the form of
\begin{eqnarray}
\label{eq:UV} H^2=\frac{16\pi}{3}
{\rho_U}\left(1-\sqrt{1-\frac{\rho}{\rho_U}}\right)\;.
\end{eqnarray}
Here $\rho$ is total energy of dark matter and radiation. Then we
obtain the Friedmann equation in GR to zero order of
${\rho}/{\rho_U}$,
\begin{eqnarray}
\label{eq:fr} 3H^2=8\pi \rho\;,
\end{eqnarray}
and the Friedmann equation in Randall-Sundrum model to the first
order of ${\rho}/{\rho_U}$ \cite{RS:1999},

\begin{eqnarray}
\label{eq:fr} 3H^2=8\pi \left(\rho+\frac{\rho^2}{2\rho_U}\right)\;.
\end{eqnarray}
It is easy to find that, at very high energy densities, the Big bang
singularity is avoided according to Eq.~(\ref{eq:UV}). The maximum
of cosmic energy density is of the order of Planck energy density
and the Universe has the minimum Hubble radius on the order of
Planck length. \emph{Thus the Universe is created from a de Sitter
phase.} We note that if $\rho_U$ is negative, the above equation
recovers to the loop quantum gravity (or extra time dimension) case
\cite{cop:2005,ash:2006,sin:2006}.
\subsection{IR modification}
In this subsection, we investigate the IR modification. We find that
the IR modification can account for the acceleration of the
Universe. Although the dark energy density contributed by this
modification behaves as phantom \cite{cal:2002}, the Big-Rip
singularity can be avoided. For the IR modification, the Friedmann
equation is given by
\begin{eqnarray}
\label{eq:fr} 3H^2=8\pi {\rho}+\frac{16\pi^2\rho_I^2}{3H^2}\;.
\end{eqnarray}
It is a quadratic equation of $H^2$. The physical solution is given by

\begin{eqnarray}
\label{eq:IR} 3H^2=4\pi
{\rho}\left(1+\sqrt{1+\frac{\rho_I^2}{\rho^2}}\right)\;.
\end{eqnarray}
Then we obtain the Friedmann equation in GR to zero order of
${\rho_I}/{\rho}$,
\begin{eqnarray}
\label{eq:fr} 3H^2=8\pi \rho\;,
\end{eqnarray}
and one Friedmann equation in ``Cardassian models" \cite{fre:2002}
to the first order of ${\rho_I}/{\rho}$

\begin{eqnarray}
\label{eq:fr} 3H^2=8\pi \left(\rho+\frac{\rho_I^2}{4\rho}\right)\;.
\end{eqnarray}
In ``Cardassian models" \cite{fre:2002}, the Friedmann equation is
modified as

\begin{eqnarray}
\label{eq:fr} 3H^2=8\pi \left(\rho+B\rho^{\eta}\right)\;,
\end{eqnarray}
with $B$ a constant. Supernova and CMB suggest $\eta\leq 0.4$ \cite{fre:2002}. It is
easy to find that the Big Rip or Big Collapse singularity is avoided
according to Eq.~(\ref{eq:IR}). With the diluting of cosmic matter,
\emph{the Universe ends in a de Sitter phase}. The minimum of cosmic
energy density is $\rho_I/2$ and the Universe has the maximum but
finite Hubble radius.

In the next, let's show the IR modification can account for the
acceleration of the Universe. For the present-day Universe, we have
\begin{eqnarray}
\label{eq:pfe} 3H_0^2=8\pi\rho_0\;,
\end{eqnarray}
where $H_0$ and $\rho_0$ are the present-day Hubble parameter and
the present-day total energy density. Divided Eq.~(\ref{eq:IR}) by
Eq.~(\ref{eq:pfe}) and put
\begin{eqnarray}
h=\frac{H}{H_0}\;,\ \ \ \Omega_{m0}=\frac{\rho_{m0}}{\rho_0}\;,\ \ \
\varepsilon=\frac{\rho_I}{\rho_{m0}}\;,
\end{eqnarray}
where $\Omega_{m0}$ is the relative density of the dark matter (For
the matter dominated Universe, we can safely neglect radiation
matter). The Friedmann equation is reduced to
\begin{eqnarray}
\label{eq:h}
{h}^2&=&\frac{1}{2}\Omega_{m0}a^{-3}\left(1+\sqrt{1+\varepsilon^2a^6}\right)\;,
\end{eqnarray}
Apply above equation on the present-day Universe ($a=1, h=1$), we
have

\begin{eqnarray}
\varepsilon=2\Omega_{m0}^{-1}\sqrt{1-\Omega_{m0}}\;.
\end{eqnarray}
The present-day matter density parameter $\Omega_{m0}$ has been
obtained by Komatsu et al. \cite{kom:2008} from a combination of
baryon acoustic oscillation, type Ia supernovae and WMAP5 data at a
$95\%$ confidence limit, $\Omega_{m0}=0.25$. So in the following
discussions, we will put $\Omega_{m0}=0.25$.

Thus same as $\Lambda \textrm{CDM}$ model, the IR model is also one
parameter model. Then ratio of dark energy density is given by
\begin{eqnarray}
\Omega_X=\frac{1}{2}\Omega_{m0}a^{-3}h^{-2}\left(-1+\sqrt{1+\varepsilon^2a^6}\right)\;.
\end{eqnarray}
 In Fig.~1 and Fig.~2, we plot the
evolution of density ratios for dark energy, dark matter and the
equation of state of dark energy. We see this dark energy model
behaves as phantom matter. The dark energy density is negligible at
the redshifts greater than $2$. Therefore the theories of structure
formation and nucleosynthesis would not be modified. Actually, we
can understand this point from Eq.~(28). At higher redshift (large H),
dark energy is negligible. At late times (small H), dark energy
becomes significant and dominant. In Fig.~3, we plot the Hubble
parameter and redshift relations for $\Lambda \textrm{CDM}$ model
and the IR model with the same parameters $\Omega_{m0}$. Both models
are very well consistent with observation data. In order to show the
IR model can account for the acceleration of the Universe, we plot
the deceleration parameter $q$
\begin{equation}
q=\frac{1}{2}+\frac{3p_X}{2\rho_X+2\rho_{m}},
\end{equation}
for $\Lambda \textrm{CDM}$ model and IR model. We find the two
models predict the same transition redshift of the Universe from
deceleration to acceleration at $z_T\simeq 0.8$.

We note that by assuming the dark energy is proportional to
$H^{\eta}$, Dvali and Turner \cite{dva:2003} have constrained
$\eta\leq 1$ with observations. Therefore, our IR modifications is
observationally viable.

\begin{figure}
\includegraphics[width=6.5cm]{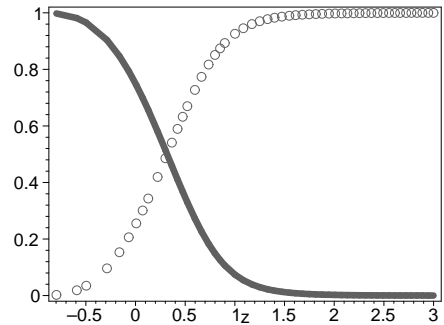}
\caption{The ratio of densities for dark matter (circled line) and dark energy (solid line).
The cosmic coincidence problem is relaxed. Here we put
$\Omega_{m0}=0.25$.}
\end{figure}

\begin{figure}
\includegraphics[width=6.5cm]{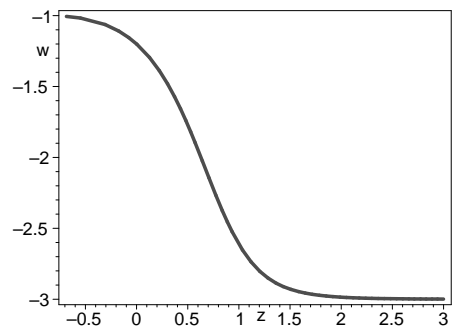}
\caption{The evolution of the equation of state for dark energy.
This is a phantom dark energy. Here we put $\Omega_{m0}=0.25$.}
\end{figure}
\begin{figure}
\includegraphics[width=6.5cm]{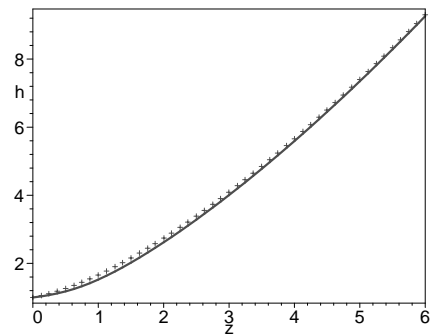}
\caption{The Hubble-redshift relations for $\Lambda \textrm{CDM}$
model (pointed line) and the IR model (solid line). Both models are
consistent with the observational data. Here we put
$\Omega_{m0}=0.25$.}
\end{figure}

\begin{figure}
\includegraphics[width=6.5cm]{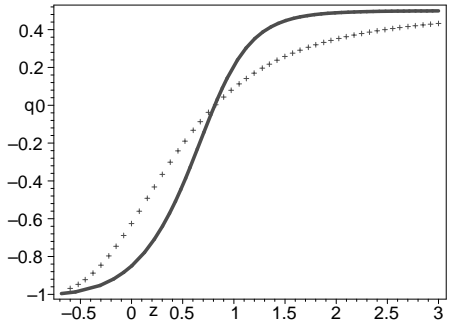}
\caption{The evolution of decelerating parameters for $\Lambda
\textrm{CDM}$ model (pointed line) and the IR model (solid line).
Both models predict the transition redshift of the Universe from
deceleration to acceleration at $z_T\simeq0.8$. Here we put
$\Omega_{m0}=0.25$.}
\end{figure}

\section{static spherically vacuum solution-f(R) kind}
In this section, we shall present the static, spherically symmetric
and vacuum solutions to verify whether it meets the solar system tests. The metric takes the form of

\begin{eqnarray}
ds^2=-N\left(r\right)^2dt^2+\frac{1}{f\left(r\right)}dr^2+r^2d\Omega^2\;.
\end{eqnarray}
We find
\begin{eqnarray}
K_{ij}=0\;,\ \ \ \ R^{(3)}=\frac{2}{r^2}\left(rf^{'}-1+f\right)\;,
\end{eqnarray}
where prime denotes the derivative with respect to $r$. So the
action for the gravitational sector can be written as
\begin{eqnarray}
\label{eq:fr}
 S=\int dtd^3x\frac{N}{16\pi\sqrt{f}}\left[R^{(3)}+\alpha \left(R^{(3)}\right)^2+\beta \left(R^{(3)}\right)^{-1}\right]\;.
\end{eqnarray}
In the first place, let's look for the solution for UV modification.
In this case, we should put $\beta=0$. Variation of action with
respect to $N$ yields
\begin{eqnarray}
\label{eq:fr} R^{(3)}+\alpha \left(R^{(3)}\right)^2=0\;.
\end{eqnarray}
Solving the equation, we obtain two solutions

\begin{eqnarray}
\label{eq:fr} f=1-\frac{2M}{r}-\frac{r^2}{6\alpha}\;,
\end{eqnarray}
and
\begin{eqnarray}
\label{eq:fr} f=1-\frac{2M}{r}\;,
\end{eqnarray}
where $M$ is an integration constant which has the meaning of the
mass of gravitational source. On the other hand, variation of action
with respect to $f$ yields

\begin{eqnarray}
\label{eq:fr} \left(r^4-6\alpha r^2+12M\alpha
r\right)N^{'}-Nr^3+6NM\alpha=0\;,
\end{eqnarray}
from which we obtain

\begin{eqnarray}
\label{eq:fr} N^2=1-\frac{2M}{r}-\frac{r^2}{6\alpha}\;,
\end{eqnarray}

and
\begin{eqnarray}
\label{eq:fr} N^2=1-\frac{2M}{r}\;.
\end{eqnarray}

Naively, the static, spherically symmetric and vacuum solution to UV
modification is the Schwarzschild or Schwarzschild-de Sitter solution. However,
it is easy to find that in the limit of $\alpha\rightarrow 0$ and $\beta\rightarrow 0$, the action of Eq.~(6) would smoothly match GR.
But this Schwarzschild-de Sitter solution would be divergent when $\alpha\rightarrow 0$. Therefore, the physical solution is uniquely left with the Schwarzschild solution.

Secondly, let's look for the solution for IR modification. In this
case, we should put $\alpha=0$. Variation of action with respect to
$N$ yields
\begin{eqnarray}
\label{eq:fr} R^{(3)}+\beta \left(R^{(3)}\right)^{-1}=0\;.
\end{eqnarray}
Solving the equation, we obtain

\begin{eqnarray}
\label{eq:fr} f=1-\frac{2M}{r}\;,\ \ \ \beta=0\;,
\end{eqnarray}
where $M$ also stands for an integration constant. On the other
hand, variation of action with respect to $f$ yields

\begin{eqnarray}
\label{eq:fr} \left(6r^2-12M
r\right)N^{'}+6NM=0\;,
\end{eqnarray}
from which we obtain

\begin{eqnarray}
\label{eq:fr} N^2=1-\frac{2M}{r}\;.
\end{eqnarray}
Therefore, the static, spherically
symmetric and vacuum solution to IR modification is exactly the
Schwarzschild solution. Since the solar system tests mainly base on the schwarzschild solution, we conclude the theory is viable for solar system tests.

\section{Cosmology-K-essence kind}
In this section, let's investigate the cosmic behavior of the modified gravity for K-essence kind.
The corresponding action is then
given by
\begin{eqnarray}
 S=\int dtd^3x\frac{Na^3}{16\pi}\left[X+{F}\left(X\right)\right]+S_{m}\;.
\end{eqnarray}
Variation of the action with respect to $N$ and then put $N=1$, we
obtain the Friedmann equation

\begin{eqnarray}
 3H^2=8\pi\sum_i\rho_i-\frac{{F}}{2}-6H^2F^{'}\;.
\end{eqnarray}
On the other hand, variation of the metric with respect to
$a\left(t\right)$ and then put $N=1$, we obtain the acceleration
equation

\begin{eqnarray}
 2\dot{H}+3H^2&=&-8\pi \sum_i
 p_i-\frac{F}{2}\nonumber\\&&-\left(2\dot{H}+6H^2\right)F^{'}+24H^2\dot{H}F^{''}\;.
\end{eqnarray}
Here $\rho_i, p_i$ are as defined before. The prime denotes the
derivative with respect to $X$. We assume there is no interaction between dark matter and
radiation. So the energy conservation equation
\begin{eqnarray}
 \frac{d\rho_i}{dt}+3H\left(\rho_i+p_i\right)=0\;,
\end{eqnarray}
still holds. For convenience we shall investigate the exponential
function for F:
\begin{eqnarray}
 F=F_0e^{\zeta X}\;,
\end{eqnarray}
where $F_0, \zeta$ are two constants. Then the Friedmann equation
is given by
\begin{eqnarray}
 3H^2=8\pi\sum_i\rho_i-F_0e^{-6\zeta H^2}\left(\frac{1}{2}+6\zeta H^2\right)\;.
\end{eqnarray}
With the usual definitions

\begin{eqnarray}
 \Omega_{i}=\frac{\rho_i}{\rho_0}\;, \ \ \ h=\frac{H}{H_0}\;,
\end{eqnarray}
the Friedmann equation becomes

\begin{eqnarray}
 h^2=\frac{\Omega_{m0}}{a^3}+\frac{\Omega_{r0}}{a^4}+f_0 e^{-\xi
 h^2}\left(\frac{1}{2}+\xi h^2\right)\;,
\end{eqnarray}

Here $\rho_0, H_0$ are the present-day total cosmic energy density
and the present-day Hubble parameter. $\Omega_{m0}, \Omega_{r0}$ are
the relative density of dark matter and radiation in present-day
Universe. We have defined:

\begin{eqnarray}
 f_0\equiv-\frac{F_0}{3H_0^2}\;\ \ \ \ \xi=6\zeta H_0^2\;.
\end{eqnarray}

Apply above equation on the present-day Universe ($a=1, h=1$), we
have

\begin{eqnarray}
f_0=\frac{2\left(1-\Omega_{m0}-\Omega_{r0}\right)}{e^{-\xi}\left(1+2\xi\right)}\;.
\end{eqnarray}
The ratio of dark energy density is given by
\begin{eqnarray}
\Omega_X=f_0 e^{-\xi
 h^2}\left(\frac{1}{2h^2}+\xi\right)\;.
\end{eqnarray}

In Fig.~5 and Fig.~6, we plot the equation of state of dark energy
for different parameters, $\xi=0.36,\
0.66,\ 1.26$ and $\xi=0.01$, respectively. We find that when $\xi<0.66$, the dark
energy model behaves as quintom matter \cite{feng:2004} which can
crosse phantom divide smoothly. On the other hand, when $\xi\geq
0.66$, the dark energy behaves as phantom matter \cite{cal:2002}
which always have the equation of state $w<-1$. When $\xi=0$, it
reduces to the cosmological constant. We see this dark energy is
negligible at the high redshifts. Therefore the theories of
structure formation and nucleosynthesis would not be modified. In
order to mimic $\Lambda \textrm{CDM}$ model at most, in the
following, we will consider $\xi=0.01$.

In Fig.~7, we plot the relative densities for radiation, dark matter
and dark energy. We see this dark energy is negligible at the high
redshifts. It is dominant only at very late time. To show the model can account for the acceleration of the Universe,
we plot the deceleration parameter $q$
\begin{equation}
q\equiv\frac{1}{2}\left(1+\frac{3p_{tot}}{\rho_{tot}}\right)\;,
\end{equation}
for our model and $\Lambda \textrm{CDM}$ model in Fig.~8.
$\rho_{tot},p_{tot}$ denote the total cosmic density and total
pressure. We find the two models predict nearly the same behavior of the Universe from deceleration to acceleration.
This is because the equation of state for dark energy is $w\simeq -1$ at the redshifts $0-2$ (see Fig.~(6)).
Therefore, the energy density of this dark energy is nearly a constant at the redshifts $0-2$.

\begin{figure}
\includegraphics[width=6.5cm]{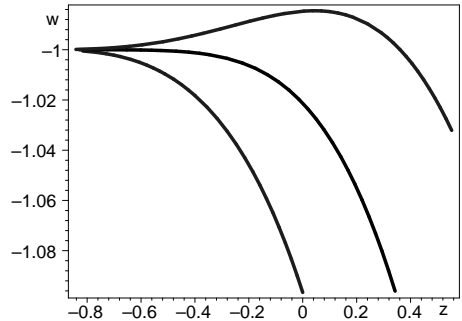}
\caption{Evolution of the equation of state for dark energy for
$\xi=0.36,\ 0.66,\ 1.26$ up down. When $\xi<0.66$, the dark
energy model behaves as quintom matter which can crosse phantom
divide smoothly. When $\xi\geq 0.66$, the dark energy behaves as
phantom matter which always have the equation of state $w<-1$. When
$\xi=0$, it reduces to the cosmological constant. Here we put
$\Omega_{m0}=0.25,\ \Omega_{r0}=8.1\cdot10^{-5}$.}
\end{figure}

\begin{figure}
\includegraphics[width=6.5cm]{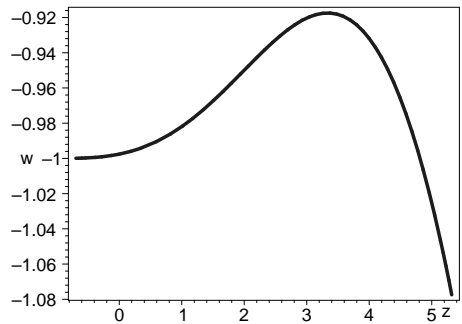}
\caption{The equation of state for a quintom dark energy model. Here
we put $\Omega_{m0}=0.25,\ \xi=0.01,\
\Omega_{r0}=8.1\cdot10^{-5}$.}
\end{figure}

\begin{figure}
\includegraphics[width=6.5cm]{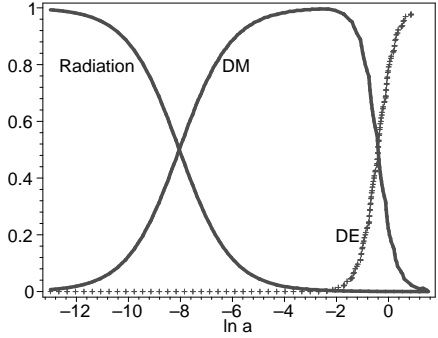}
\caption{Relative densities for radiation (solid line), dark matter
(DM) (solid line) and dark energy (DE) (dotted line). Here we put
$\Omega_{m0}=0.25,\ \xi=0.01,\ \Omega_{r0}=8.1\cdot10^{-5}$.}
\end{figure}

\begin{figure}
\includegraphics[width=6.5cm]{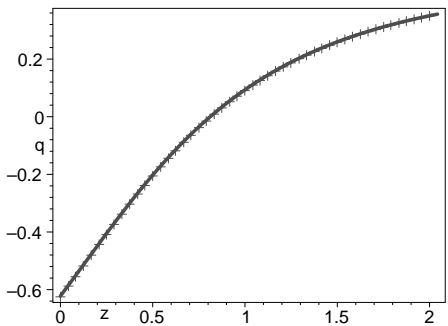}
\caption{Evolution of decelerating parameters for $\Lambda
\textrm{CDM}$ model (crossed line) and our model (solid line). Both
models predict nearly the same transition redshift of the Universe
from deceleration to acceleration at $z_T=0.8$. Here we put
$\Omega_{m0}=0.25,\ \xi=0.01,\ \Omega_{r0}=8.1\cdot10^{-5}$.}
\end{figure}

\section{static, spherically and vacuum solution-K-essence kind}
In this section, we shall present the static, spherically symmetric
and vacuum solution. The general form for a metric describing the
static, spherically symmetric spacetime is given by

\begin{eqnarray}
ds^2=-N\left(r\right)^2dt^2+\frac{1}{f\left(r\right)}dr^2+r^2d\Omega^2\;.
\end{eqnarray}
Using the metric, we find the extrinsic curvature and the three
dimensional Ricci scalar are
\begin{eqnarray}
K_{ij}=0\;,\ \ \ \ R=-\frac{2}{r^2}\left(rf^{'}-1+f\right)\;,
\end{eqnarray}
where prime denotes the derivative with respect to $r$. So the
action for the gravitational sector can be written as
\begin{eqnarray}
 S=\int dtd^3x\frac{N}{16\pi\sqrt{f}}\left[-\frac{2}{r^2}\left(rf^{'}-1+f\right)+F_0\right]\;.
\end{eqnarray}

Variation of action with respect to $N$ yields
\begin{eqnarray}
\frac{2}{r^2}\left(rf^{'}-1+f\right)-F_0=0\;.
\end{eqnarray}
Solving the equation, we obtain

\begin{eqnarray}
f=1-\frac{2M}{r}+\frac{F_0}{6}r^2\;,
\end{eqnarray}
where $M$ is an integration constant which has the meaning of the
mass of gravitational source. On the other hand, variation of action
with respect to $f$ yields

\begin{eqnarray}
\left(-F_0r^4+12M
r-6r^2\right)N^{'}+\left(6M+F_0r^3\right)N=0\;,
\end{eqnarray}
from which we obtain

\begin{eqnarray}
N^2=f&=&1-\frac{2M}{r}+\frac{F_0}{6}r^2\nonumber\\
&=&1-\frac{2M}{r}-\frac{f_0}{2}H_0^2r^2\;.
\end{eqnarray}
Equation (61) tells us the dimensionless constant $f_0\simeq 1.48$
for $\Omega_{m0}=0.25,\ \xi=0.01,\ \Omega_{r0}=8.1\cdot 10^{-5}$.
So the static, spherically symmetric and vacuum solution is the Schwarzschild-de Sitter solution.
The solar system tests constrain the Schwarzschild-de Sitter metric that $H_0^2<10^{-41}\textrm{m}^{-2}$ (see, e.g. \cite{val:2006}).
Take the present-day Hubble parameter as $H_0=71\ \textrm{km}\ \textrm{sec}^{-1}\ \textrm{Mpc}^{-1}$,  we then obtain $H_0^2=6.6\cdot 10^{-57}\textrm{m}^{-2}$.
Therefore, the theory is not conflict with solar system tests.
\section{conclusion and discussion}
In conclusion, we have investigated two kinds of modified gravity theories in
ADM formalism. The Friedmann equation of $f(R)$ kind is remarkably
simple and very different from that in the Jordan frame.
The UV modification can avoid the Big-Bang
singularity and the IR modification can avoid the Big-Rip
singularity, respectively. In this version, the Universe starts from a
de Sitter phase and ends in another de Sitter phase. For the K-essence modified gravity, the Universe starts from Big-Bang but ends
in de Sitter phase. It is interesting that the corresponding dark energy behaves as quintom matter. We find both theories can account for the current acceleration of the Universe without invoking any dark energy.

We also find the static, spherically symmetry and vacuum solutions
to both theories. The solutions are the Schwarzschild or Schwarzschild-de Sitter solution. We verify that
the solutions are viable for solar system tests. In view of above simple and interesting results, the modified gravities in the ADM formalism merit further detailed study.

\acknowledgments

I thank the anonymous referee for the insightful
 comments and suggestions, which have allowed me to improve this paper significantly. I also thank Rong-Gen Cai, Xin Zhang, Hao Wei for stimulating and illuminating discussions. Special thanks go to Xiaoning Wu, Yu Tian, Sijie Gao and Yun-Song Piao for their comments which greatly complete the paper. This work is
supported by the National Science Foundation of China under the
Distinguished Young Scholar Grant 10525314, the Key Project Grant
10533010, Grant 10575004, Grant
10973014 and the 973 Project.

\newcommand\AL[3]{~Astron. Lett.{\bf ~#1}, #2~ (#3)}
\newcommand\AP[3]{~Astropart. Phys.{\bf ~#1}, #2~ (#3)}
\newcommand\AJ[3]{~Astron. J.{\bf ~#1}, #2~(#3)}
\newcommand\APJ[3]{~Astrophys. J.{\bf ~#1}, #2~ (#3)}
\newcommand\APJL[3]{~Astrophys. J. Lett. {\bf ~#1}, L#2~(#3)}
\newcommand\APJS[3]{~Astrophys. J. Suppl. Ser.{\bf ~#1}, #2~(#3)}
\newcommand\JCAP[3]{~JCAP. {\bf ~#1}, #2~ (#3)}
\newcommand\LRR[3]{~Living Rev. Relativity. {\bf ~#1}, #2~ (#3)}
\newcommand\MNRAS[3]{~Mon. Not. R. Astron. Soc.{\bf ~#1}, #2~(#3)}
\newcommand\MNRASL[3]{~Mon. Not. R. Astron. Soc.{\bf ~#1}, L#2~(#3)}
\newcommand\NPB[3]{~Nucl. Phys. B{\bf ~#1}, #2~(#3)}
\newcommand\PLB[3]{~Phys. Lett. B{\bf ~#1}, #2~(#3)}
\newcommand\PRL[3]{~Phys. Rev. Lett.{\bf ~#1}, #2~(#3)}
\newcommand\PR[3]{~Phys. Rep.{\bf ~#1}, #2~(#3)}
\newcommand\PRD[3]{~Phys. Rev. D{\bf ~#1}, #2~(#3)}
\newcommand\RMP[3]{~Rev. Mod. Phys.{\bf ~#1}, #2~(#3)}
\newcommand\SJNP[3]{~Sov. J. Nucl. Phys.{\bf ~#1}, #2~(#3)}
\newcommand\ZPC[3]{~Z. Phys. C{\bf ~#1}, #2~(#3)}


\begin{thebibliography}{99}

\bibitem{ho:20081}P. Horava, arXiv:0811.2217 [hep-th].
\bibitem{ho:20082} P. Horava, JHEP 0903, 020 (2009).
\bibitem{ho:20083}P. Horava, Phys. Rev. D 79, 084008 (2009).
\bibitem{ho:20090} P. Horava,  arXiv:0902.3657 [hep-th].
\bibitem{ho:20091} T.Takahashi and J. Soda,  arXiv:0904.0554 [hep-th].
\bibitem{ho:20092} G. Calcagni,
 arXiv:0904.0829 [hep-th].
\bibitem{ho:20093}E. Kiritsis and G. Kofinas,
arXiv:0904.1334 [hep-th].
\bibitem{ho:20094}J. Kluson,  arXiv:0904.1343
[hep-th].
\bibitem{ho:20095}H. Lu, J. Mei and C. N. Pope, arXiv:0904.1595
[hep-th].
\bibitem{ho:20096}S. Mukohyama,  arXiv:0904.2190
[hep-th].
\bibitem{ho:20097}R. Brandenberger,arXiv:0904.2835 [hep-th].
\bibitem{ho:20098}H. Nikolic, arXiv:0904.3412 [hep-th]. 11.
\bibitem{ho:20099} H.Nastase, arXiv:0904.3604 [hepth].
\bibitem{ho:200910}R. G. Cai, L. M. Cao and N. Ohta, arXiv:0904.3670
[hep-th].
\bibitem{ho:200911}X. Gao, arXiv:0904.4187 [hep-th].
\bibitem{ho:200912}B. Chen, Q.G. Huang, arXiv:0904.4565 (hep-th);
\bibitem{ho:200914}T. P. Sotiriou, M. Visser, S. Weinfurtner,
arXiv:0904.4464 (hep-th).
\bibitem{ho:200915} Eoin o Colgain, Hossein Yavartanoo, arXiv:0904.4357
\bibitem{ho:200916} Y. S. Piao, arXiv:0904.4117 (hep-th)
\bibitem{ho:200917} G.E. Volovik, arXiv:0904.4113 (hep-ph)
\bibitem{ho:200918}R. G. Cai, Y. Liu, Y. W. Sun, arXiv:0904.4104 (hep-th)


\bibitem{adm:1962} R.L. Arnowitt, S. Deser and C.W. Misner, The dynamics of general
relativity, ¡°Gravitation: an introduction to current research¡±,
Louis Witten ed. (Wilew 1962), chapter 7, pp 227-265,
arXiv:gr-qc/0405109.

\bibitem{st:1977}S. Capozziello and S. Tsujikawa, \PRD{77}{107501}{2008}; S. M. Carroll, V. Duvvuri, M. Trodden, M. S. Turner, \PRD{70}{043528}{2004}
; Strobinsky A 1977, JETP Lett. 30, 682;
Capzelo S,O hionero F and Amendola L 1993, Int. J Mod. Phys. D1,615;
Amendola L, Litterio M and Occhionero, 1990, Int. J Mod. Phys A
5,3861; Stel K S 1977, Phys. Rev. D 16, 953; Buchbinder I L,
Odintsov S D and Shapiro I L, 1992, Effective Action in Quantum
Gravity (Bristol: IOP); S. Nojiri and S.D.Ordintsov, Phys.Rev.D 68,
123512(2003); S. Nojiri and S.D.Ordintsov, Int. J. Geom. Meth. Mod.
Phys.4, 115-146, 2007.
\bibitem{wald}C. Misner, K. S. Thorn and J. Wheeler (1973), Gravitation, (San Francisco: W. H. Freeman and
Company), Chapter 21, Page 519.

\bibitem{k-essence:00}C. Armendariz-Picon, V. F. Mukhanov and P. J. Steinhardt, Phys. Rev. Lett. 85, 4438 (2000)
[astro-ph/0004134]; C. Armendariz-Picon, V. F. Mukhanov and P. J.
Steinhardt, Phys. Rev. D 63, 103510 (2001) [astro-ph/0006373].
\bibitem{bin:2000}P. Binetruy, C. Deffayet, and D. Langlois, Nucl. Phys. B565, 269 (2000)
(hep-th/9905012); P. Binetruy, C. Deffayet, U. Ellwanger, and D.
Langlois, Phys. Lett. B477, 285 (2000) (hep-th/9910219).
\bibitem{dva:2002}G.Dvali,G.Gabadadze,andM.Shifman,hep-th/0202174.
\bibitem{dva:2003}G.Dvali and M.S.Turner,astr-ph/0301510.
\bibitem{RS:1999}L. Randall and R. Sundrum, Phys. Rev. Lett. 83, 4690 (1999)
[arXiv:hep-th/9906064]; P. Binetruy, C. Deffayet, U. Ellwanger and
D. Langlois, Phys. Lett. B 477, 285 (2000) [arXiv:hep-th/9910219].

\bibitem{cop:2005} E. J. Copeland, S.-J. Lee, J. E. Lidsey and S. Mizuno, Phys. Rev. D 71, 023526 (2005) [astroph/
0410110].

\bibitem{ash:2006}A. Ashtekar, T. Pawlowski and P. Singh, Phys. Rev. Lett. 96, 141301
(2006) [arXiv:gr-qc/0602086]. A. Ashtekar, T. Pawlowski and P.
Singh, Phys. Rev. D 73, 124038 (2006) [arXiv:gr-qc/0604013]. A.
Ashtekar, T. Pawlowski and P. Singh, Phys. Rev. D 74, 084003 (2006)
[arXiv:gr-qc/0607039]. A. Ashtekar, T. Pawlowski, P. Singh and K.
Vandersloot, Phys. Rev. D 75, 024035 (2007) [arXiv:gr-qc/0612104].
K. Vandersloot, Phys. Rev. D 75, 023523 (2007)
[arXiv:gr-qc/0612070].

\bibitem{sin:2006} P. Singh, K. Vandersloot
and G. V. Vereshchagin, Phys. Rev. D 74, 043510 (2006)
[arXiv:gr-qc/0606032].

\bibitem{feng:2004}B. Feng, X. L. Wang and X. M. Zhang, Phys. Lett. B 607, 35 (2005) [astro-ph/0404224];
B. Feng, M. Li, Y. S. Piao and X. M. Zhang, Phys. Lett. B 634, 101
(2006) [astro-ph/0407432];
\bibitem{cal:2002}R.R.Caldwell, Phys. Lett. B545, 23 (2002).
\bibitem{fre:2002}K. Freese and M. Lewis, Phys. Lett. B 540, 1 (2002)
[arXiv:astro-ph/0201229].
\bibitem{kom:2008}E. Komatsu et al., 2008, arXiv:0803.0547.

\bibitem{val:2006} V. Kagramanova, J. Kunz, C. Lmmerzahl, \PLB{634}{465}{2006}.

























\end{thebibliography}
\end{document}